\documentclass[12pt]{article}
\usepackage{epsfig}
\usepackage{amsfonts}
\usepackage{amscd}
\usepackage{latexsym}
\usepackage{amsmath,amssymb}
\usepackage{verbatim}
\usepackage{setspace}
\usepackage{color}
\usepackage{cite}

\usepackage[textheight=9in, textwidth=6.5in, letterpaper]{geometry}

\usepackage{color}   %May be necessary if you want to color links
\usepackage{hyperref}
\hypersetup{
    colorlinks=true,  %set true if you want colored links
    linktoc=all,     %set to all if you want both sections and subsections linked
    linkcolor=black,  %choose some color if you want links to stand out
    citecolor=black,
    filecolor=black,
    urlcolor=black,
}

\numberwithin{equation}{section}

\def\p{\partial}

\def\<{\langle}
\def\>{\rangle}

\def\cD{{\cal  D}}

\def\cO{\mathcal{O}}

\def\ip{{\cal I}^+}
\def\im{{\cal I}^-}

\def\be{\begin{equation}}
\def\ee{\end{equation}}
\def\beq{\be\begin{array}{c}}
\def\eeq{\end{array}\ee}
\def\bes{\be\begin{split}}
\def\ees{\end{split} \ee}
\def\bs{\begin{split}}
\def\es{\end{split} }
\def\nn{\nonumber}

\def\cs{\mathcal{S}}

\def\b{{\beta}}
\def\a{{\alpha}}

\def\e{{\epsilon}}

   \makeatletter
  \let\over=\@@over \let\overwithdelims=\@@overwithdelims
  \let\atop=\@@atop \let\atopwithdelims=\@@atopwithdelims
  \let\above=\@@above \let\abovewithdelims=\@@abovewithdelims
\renewcommand\section{\@startsection {section}{1}{\z@}%
                                   {-3.5ex \@plus -1ex \@minus -.2ex}%nn
                                   {2.3ex \@plus.2ex}%
                                   {\normalfont\large\bfseries}}

\renewcommand\subsection{\@startsection{subsection}{2}{\z@}%
                                     {-3.25ex\@plus -1ex \@minus -.2ex}%
                                     {1.5ex \@plus .2ex}%
                                     {\normalfont\bfseries}}

\begin{document}
\begin{titlepage}
\unitlength = 1mm
{\begin{flushright} CALT-TH-2015-060\end{flushright}}
\ \\
\vskip 1cm
\begin{center}

{ \LARGE {\textsc{Asymptotic Fermionic Symmetry From Soft  Gravitino Theorem }}}

\vspace{0.8cm}
 Vyacheslav Lysov

\vspace{1cm}

{\it Walter Burke Institute for Theoretical Physics,\\
 California Institute of Technology, \\
 Pasadena, CA 91125, USA}

\begin{abstract}
We discuss the semiclassical scattering problem for  massless matter coupled to Rarita-Schwinger field in four dimensional Minkowski space.
We rewrite the soft gravitino  theorem as a Ward identity for the $\mathcal{S}$-matrix and  discuss the relationship of    corresponding symmetry charges
to  new asymptotic fermonic symmetries of massless Rarita-Schwinger field.

\end{abstract}%\vspace{0.5cm}

\vspace{1.0cm}

\end{center}

\end{titlepage}

\pagestyle{empty}
\pagestyle{plain}

\pagenumbering{arabic}

\tableofcontents
\section{Introduction}

Our work was inspired by the recent progress in understanding soft theorems as Ward identities for certain asymptotic symmetries. The possibility for such correspondence can be guessed from an observation that the points on null infinity are causally disconnected.  We can define conserved charges for each generator of null infinity,  which  is essentially  the same as introducing continuous symmetry parameters (functions, vector fields or  spinors on  two dimensional sphere at infinity) for global symmetries in four dimensional Minkowski. Symmetries act on massless scattering data   at asymptotic infinity, which may include gravitational data. Gravitational scattering data is the same as geometry, so the symmetry transform one asymptotically Minkowski space into another asymptotically Minkowski space, which makes this symmetry {\it asymptotical} in contrast with ordinary symmetry that preserves particular metric.  In pure gravity such symmetry was discovered by Bondi, Van der Berg, Metzner and Sachs \cite{Bondi:1962px,Sachs:1962wk,Sachs:1962zza} and is often denoted as the BMS symmetry.

It is not surprising that the asymptotic symmetry is a symmetry of the massless $\cs$-matrix, however  the precise relation requires careful description  of the  scattering data and  matching across spatial infinity. 
Furthermore, the Ward identity for an asymptotic symmetry is nothing but a soft theorem,  discovered  in context of scattering theory by Weinberg \cite{Weinberg:1965nx}.   The relationship between the asymptotic symmetry of abelian gauge theory  and Weinberg's soft photon   theorem  was described  by Strominger \cite{Strominger:2013lka}  and was later  generalized in  \cite{Strominger:2013jfa, He:2014laa,  Kapec:2014opa,He:2014cra, Lysov:2014csa, Kapec:2014zla,Mohd:2014oja, Larkoski:2014bxa,Campiglia:2014yka,Campiglia:2015yka, Banks:2014iha,Adamo:2014yya,Geyer:2014lca,Bern:2014vva,
Kapec:2015vwa,He:2015zea,Kapec:2015ena,Avery:2015gxa,Strominger:2015bla, Campiglia:2015kxa,Avery:2015rga}. Moreover it has been shown \cite{Strominger:2014pwa,Pasterski:2015tva,Pasterski:2015zua} that the changes in asymptotic data   can be measured in a form of   various memory effects.

Most of  the asymptotic symmetries, that were successfully related to  soft theorems, were bosonic, except the recent work \cite{Dumitrescu:2015fej}, where authors discovered an infinite-dimensional
sermonic symmetry in supersymmetric QED coupled to massless  charged matter. 

Supersymmetry algebra of Minkowski space  includes Poincare transformations as bosonic generators, which asymptotically are elevated to the infinite-dimensional BMS algebra \cite{Bondi:1962px,Sachs:1962wk,Sachs:1962zza,Barnich}, so it in natural to address a question whether such an infinite-dimensional generalization exists for global supersymmetries.    
Apparently such  generalization was discussed by  Awada, Gibbons and Shaw\cite{Awada:1985by}, while  the corresponding soft theorems
were studied by Grisaru and Pendleton  \cite{Grisaru} for pure $\mathcal{N}=1$ supergravity case  and further expanded using modern spinor-helicity formalism to $\mathcal{N}=8$ case by \cite{Liu:2014vva}.

In this article, we will  use  the simplest $\mathcal{N}=1$ soft gravitino  theorem, recast it  into a Ward, identity and analyze the symmetry  generated by this procedure. It turns out that this symmetry, parametrized by a spinor on two sphere,  generates a  local supersymmetry  transformation  on massless supermatter and changes gravitino by a zero mode, so it can be identified with a sermonic asymptotic symmetry.

The paper is organized as follows: in section 2 we introduce a Rarita-Shwinger field coupled to supersymmetric matter. We describe gauge fixing, coordinate system   in vicinity of asymptotic null infinity and define  4d spinor decomposition in terms of the 2d spinors.  In section 3  we describe  free scattering data on null infinity and relate it to the  plane wave expansion and zero modes.   In section 4 we derive a Ward identity from the gravitino's soft theorem and identify symmetry charges. We discuss symmetry transformations generated by these charges in section 5 and propose their relation to asymptotic symmetries.

\section{Rarita-Schwinger field}
Massless   gravitino field $\psi_\mu$ coupled to supersymmetric matter is described by Rarita-Schwinger (RS)  equation  
\be\label{RSM}
\gamma^{\mu\nu\rho} \nabla_\nu \psi_{\rho}=J^\mu,
\ee
where $\gamma^\mu$ are 4d gamma matrices,  $J^\mu$ is a supercurrent for the matter fields.  Note that covariant derivative $\nabla_\mu$ may have a nontrivial spin-connection part. Equation (\ref{RSM}) is invariant under gauge transformations 
\be\label{gauge_transform}
\delta \psi_\mu = \nabla_\mu \e 
\ee
where $\e$ is a four dimensional spinor.   Gauge transformations (\ref{gauge_transform}) are local supersymmetries in purely bosonic background. 
In order to describe the radiation data for RS equation we need to fix a gauge. We are going to analyze the (\ref{RSM}) equation in four dimensional Minkowski so will use the most common covariant gauge 
\be\label{gauge_fix}
\nabla^\mu \psi_\mu = \gamma^\mu \psi_\mu =0.
\ee
These gauge conditions require that the matter supercurrent $J^\mu$ is $\gamma$-traceless
\be\label{notrace}
\gamma^\mu \cdot J_\mu=0.
\ee
Superconformal matter automatically satisfy this condition since it related to the trace of stress tensor by supersymmetry. For general supersymmetric matter in flat space  we often can an the improvement term
$\delta J^\mu = \p_\nu B^{[\mu\nu]}$ to make $\gamma$-traceless supercurrent.

Using $\nabla_\mu \gamma_\nu =0 $ and gauge fixing conditions (\ref{gauge_fix})  we can simplify RS equation 
\be
\gamma^{\mu\nu\rho} \nabla_\nu \psi_{\rho}=  \gamma^\nu \nabla_\nu \psi^\mu      = J^\mu.
\ee

Radiation data for massless fields lives at null asymptotic infinity $\mathcal{I}$ so we are going to introduce retarded coordinates  $(u,r,z^a)$ related to Cartesian coordinates $(t=x^0,x^i)$ by 
\be
u  = t-r,\;\;\; r^2 = x^i x_i,\;\;\; x^i =r\hat{x}^i(z),
\ee
where $\hat{x}^i(z^a)$ defines an embedding of unit round $S^2$ with coordinates $z^1,z^2$ in $\mathbb{R}^3$ with coordinates $x^1,x^2,x^3$.  Minkowski metric in this coordinates 
\be
ds^2 = -du^2 -2dudr +r^2 \gamma_{ab} dz^adz^b.
\ee
Here $\gamma_{ab}$ is a unit metric on the round $S^2$.  Future null infinity $\ip$ is given by the null hypersurface $(r=\infty, u, z^a)$, 
with future $(u=\infty)$ and past $(u=-\infty)$ boundaries denoted by $\ip_+$ and $\ip_-$, respectively. 

In the vicinity of the past null infinity $\im$,  we are going to introduce advanced coordinates  $(v,r,z^a)$ related to Cartesian coordinates $(t=x^0,x^i)$ by 
\be
v  = t+r,\;\;\; r^2 = x^i x_i,\;\;\; x^i =-r\hat{x}^i(z),
\ee
where $\hat{x}^i(z^a)$ is the same  embedding of unit round $S^2$ in $\mathbb{R}^3$ as before. Note in particular that the angular coordinate $z^a$ on $\mathcal{I}^-$ is antipodally related to the angular coordinate on $\mathcal{I}^+$, so that null generators of $\mathcal{I}$ passing through spatial infinity ($i^0$)  are labeled by the same numerical value of $z^a$. $\mathcal{I}^-$ is the $(r=\infty,v,z^a)$ null hypersurface, with future $(v=\infty)$ and past $(v=-\infty)$ boundaries denoted by $\mathcal{I}^-_+$ and $\mathcal{I}^-_-$, respectively. Minkowski metric in advanced coordinates takes the form
\be\label{adv_coord}
ds^2 = -dv^2+2dvdr +r^2 \gamma_{ab} dz^adz^b.
\ee
Note that most equations in advanced coordinates can be obtained from the ones in retarded coordinates by a  simple redefinition $u\to -v$ ,  which serves as a good reason to use   retarded coordinates in our detailed considerations, while presenting final statements for advanced coordinates  when needed.  

In order to define a covariant derivative $\nabla_\mu$ for spinor fields  in retarded coordinates we need to specify a spinor frame. We will use Minkowski space in Cartesian coordinates $(x^0, x^i)$ as a flat spinor frame
\be
e^0_\mu dx^\mu \equiv d x^0 = du +dr, \;\;
e^i_\mu dx^\mu \equiv  d x^i  = \hat{x}^i dr + r \p_a \hat{x}^i dz^a.
\ee
Spin connection  is trivial for such frame since one-forms $e^0_\mu dx^\mu, e^i_\mu dx^\mu$ are exact. Gamma matrices in retarded coordinates are of the form 
\beq\label{gamma_ret}
\gamma_r \equiv e_r^i \gamma_i +e_r^0 \gamma_0 = \gamma_0 + \hat{x}^i \gamma_i, \\
 \gamma_u  \equiv e_u^i \gamma_i +e_u^0 \gamma_0 = \gamma_0, \;\;\; \gamma_a  \equiv e_a^i \gamma_i +e_a^0 \gamma_0  = r \p_a\hat{x}^i \gamma_i.
\eeq
In retarded coordinates the RS equation (\ref{RSM}) takes the form
\beq\label{RSM_ret}
\gamma^r \p_r \psi_u- \gamma_r \p_u \psi_u+ r^{-1} \hat{\gamma}^a \p_a \psi_u    = J_u, \\
\gamma^r \p_r \psi_r- \gamma_r \p_u \psi_r + r^{-1} \hat{\gamma}^a \p_a \psi_r - r^{-2} \hat{\gamma}^a \psi_a   = J_r, \\
       \gamma^r \p_r \psi_a - \gamma_r \p_u \psi_a + r^{-1}\hat{\gamma}^b \cD_b \psi_a +\hat{\gamma}_a  (\psi_r-\psi_u)     = J_a,
\eeq
where we introduced  matrices $\hat{\gamma}^a\equiv r^{-1}\gamma^a$ which are indeed  gamma matrices on $S^2$.  Here $\cD_a$
is a an ordinary covariant (with respect to $\gamma_{ab}$ metric) derivative $D_a$ on the sphere with  spin connection
\be
\cD_a  = D_a +\frac12 \gamma^r \hat{\gamma}_a,
\ee

We can further simplify our analysis of the RS equations  (\ref{RSM_ret}) using decomposition of  4d spinors into pair of 2d spinors. Such decomposition is realized using pair of projectors
 \beq
P_0 = -\frac12 \gamma_r \gamma_u=-\frac12 \gamma^u \gamma^r, \;\;\;\;  P_1=-\frac12 \gamma_u \gamma_r =- \frac12 \gamma^r \gamma^u, \\
 P_0 +P_1=1,\;\;\; P_0P_1=P_1P_0=0,\;\;\; P_0P_0=P_0 ,\;\;\;\; P_1 P_1=P_1,
 \eeq
which obey the following useful properties
\be
\cD_a P_0 = P_0 \cD_a,\;\; P_0 \gamma^r = \gamma^r P_1,\;\;\;   P_1 \gamma^r = \gamma^r P_0,\;\; P_0 \gamma_r = \gamma_r P_1,\;\;\;   P_1 \gamma_r = \gamma_r P_0.
\ee
In order to analyze RS equations (\ref{RSM_ret}) near $\ip$ we assume an asymptotic expansion for gravitino field and supercurrent 
\be
\psi_\mu(r,u,z) = \sum_{n=0} \frac{1}{r^n}\psi^{(n)}_\mu (u,z),\;\;\; J_\mu(r,u,z) = \sum_{n=0} \frac{1}{r^n}J^{(n)}_\mu (u,z),
\ee
while our equations  (\ref{RSM_ret}) assume the following form 
\beq\label{RS_large_r}
(1-n)  \gamma^r \psi^{(n)}_{1u}  -\gamma_r \p_u \psi^{(n+1)}_{1u}  +  \hat{\gamma}^{a}  \cD_a \psi^{(n)}_{0u} =J^{(n+1)}_{0u}, \\
(1-n)   \gamma^r \psi^{(n)}_{0u}   -   \hat{\gamma}^{a}  \cD_a \psi^{(n)}_{1u} = J_{1u}^{(n+1)} , \\
(1-n)\gamma^r \psi^{(n)}_{1r}- \gamma_r \p_u \psi^{(n+1)}_{1r} + \hat{\gamma}^a \cD_a \psi^{(n)}_{0r}  - \hat{\gamma}^a \psi^{(n-1)}_{0a}   = J^{(n+1)}_{0r}, \\

(1-n)\gamma^r \psi^{(n)}_{0r} + \hat{\gamma}^a \cD_a \psi^{(n)}_{1r} -  \hat{\gamma}^a \psi^{(n-1)}_{1a}   = J^{(n+1)}_{1r},\\

-n\gamma^r  \psi^{(n)}_{1a}- \gamma_r \p_u \psi^{(n+1)}_{1a} +\hat{\gamma}^b  \cD_b \psi^{(n)}_{0a}  +\hat{\gamma}_a  (\psi^{(n+1)}_{0r}-\psi^{(n+1)}_{0u})     = J^{(n+1)}_{0a}, \\
       
-n\gamma^r  \psi^{(n)}_{0a} +\hat{\gamma}^b  \cD_b \psi^{(n)}_{1a}   +\hat{\gamma}_a  (\psi^{(n+1)}_{1r}-\psi^{(n+1)}_{1u})     = J^{(n+1)}_{1a}.
\eeq
The large-r falloff for supercurrent components is $J_a, J_u =\cO(r^{-2}),\; J_r=\cO(r^{-3})$, while trace free condition (\ref{notrace}) further restricts $J^{(2)}_{1u} =0$.
The gauge fixing conditions (\ref{gauge_fix}) 
\beq\label{gauge_large_r}
\gamma^r  \psi^{(n+1)}_{1r}- \gamma_r  \psi^{(n+1)}_{1u}  +  \hat{\gamma}^a  \psi^{(n)}_{0a} =0, \\
 \gamma^r \psi^{(n+1)}_{0r} + \hat{\gamma}^a \psi^{(n)}_{1a} =0, \\
(1-n) (\psi^{(n+1)}_{0r}-\psi^{(n+1)}_{0u}) - \p_u  \psi^{(n+2)}_{0r}  +  \cD^a \psi^{(n)}_{0a} - \frac1 {2} \gamma^r \hat{\gamma}^a \psi^{(n)}_{1a}   =0, \\
(1-n)(\psi^{(n+1)}_{1r}-\psi^{(n+1)}_{1u}) - \p_u  \psi^{(n+2)}_{1r}  + \cD^a \psi^{(n)}_{1a} - \frac1{2} \gamma^r \hat{\gamma}^a \psi^{(n)}_{0a} =0.  \\
\eeq
Let us note that our gauge fixing (\ref{gauge_fix}) allows for a residual gauge symmetry in the form of free solution to Dirac equation.  In four dimensional Minkowski free solution to Dirac equation 
$\e = r^{-1}\e^{(1)}_{0}(u,z)$,  so we can  further set 
 \be\label{residual_gauge}
  \psi_{0r}^{(2)}(u,z)=0.
 \ee

\section{The semiclassical scattering problem}

\subsection{Scattering data for gravitino  at $\ip$}
In this subsection we will present some details on the analysis of  RS equations  (\ref{RS_large_r}) and gauge fixing conditions  (\ref{gauge_large_r}).
The $\cO(r^{0})$ trace free condition (\ref{gauge_large_r})  
\be\label{gammatracefree}
\gamma^r  \psi^{(0)}_{1r}- \gamma_r  \psi^{(0)}_{1u}   =0,  \;\;
 \gamma^r \psi^{(0)}_{0r}=0.
\ee
Since $\gamma^r$ is invertible due to the relation $\gamma^r \gamma^r=1$ we can immediately write 
\be
\psi^{(0)}_{0r}=0.
\ee
Now let us look at  $\cO(r^{-1}, r^{-2})$ orders of  r-component of RS equation (\ref{RS_large_r}) 
\beq
- \gamma_r \p_u \psi^{(0)}_{1r}    = J^{(0)}_{0r}=0, \\
\gamma^r \psi^{(0)}_{1r}- \gamma_r \p_u \psi^{(1)}_{1r} + \hat{\gamma}^a \cD_a \psi^{(0)}_{0r}    = J^{(1)}_{0r}=0, \\
\gamma^r \psi^{(0)}_{0r} + \hat{\gamma}^a \cD_a \psi^{(0)}_{1r} = J^{(1)}_{1r}=0. 
\eeq
The last equation combined with  $\psi^{(0)}_{0r}=0$ leads to the 
\be
 \hat{\gamma}^a \cD_a \psi^{(0)}_{1r} =0.
\ee
There are no nontrivial solution to the Dirac equation on $S^{2}$
\be\label{Dirac_0}
\hat{\gamma}^a \cD_a \psi=0,\;\;\; \Rightarrow \psi=0.
\ee 
In order to prove  (\ref{Dirac_0}) let  us consider  
\be
(\hat{\gamma}^a \cD_a)^2 \psi =\hat{\gamma}^a \hat{\gamma}^b \cD_a \cD_b \psi =
(\hat{\gamma}^{[ab]} + \gamma^{ab})  \cD_a \cD_b \psi  =  ( \cD^2 +\frac14  \hat{\gamma}_{[ab]} \hat{\gamma}^{[ab]} ) \psi =\left( \cD^2-  \frac12\right) \psi,
\ee
where we used 
\be\label{covariant_com}
{[\cD_a, \cD_b]}\psi =\frac12  \hat{\gamma}_{[ab]}\psi =\frac14  (\hat{\gamma}_{a}  \hat{\gamma}_{b} -\hat{\gamma}_{b} \hat{\gamma}_{a}   )\psi.
\ee
Eigenvalues of $\cD^2$ are negative on $S^2$ since it is a compact manifold,  therefore  all nontrivial  eigenmodes of $(\hat{\gamma}^a \cD_a)^2 $ have negative eigenvalues, 
what concludes our   proof of  (\ref{Dirac_0}).

Using (\ref{Dirac_0}) and the  first equation from (\ref{gammatracefree}) we can further evaluate 
\be
 \psi^{(0)}_{0r}=\psi^{(0)}_{1r}= \psi^{(0)}_{1u}=0.
\ee
Now using  $\cO(r^{0}, r^{-1}, r^{-2})$ orders of the $u$-component RS equation  (\ref{RS_large_r}) 
\beq
  -\gamma_r \p_u \psi^{(0)}_{1u}  =J^{(0)}_{0u}=0, \\
 \gamma^r \psi^{(0)}_{1u}  -\gamma_r \p_u \psi^{(1)}_{1u}  +  \hat{\gamma}^{a}  \cD_a \psi^{(0)}_{0u} =J^{(1)}_{0u}=0,\\
  \gamma^r \psi^{(0)}_{0u}   -   \hat{\gamma}^{a}  \cD_a \psi^{(0)}_{1u} = J_{1u}^{(1)}=0,\\
-  \hat{\gamma}^{a}  \cD_a \psi^{(1)}_{1u} = J_{1u}^{(2)}=0,
\eeq
we can solve for 
\be
 \psi^{(0)}_{0r}=\psi^{(0)}_{1r}= \psi^{(0)}_{1u}=\psi^{(0)}_{0u}=\psi^{(1)}_{1u}=0.
\ee
Similar analysis of the system (\ref{RS_large_r}, \ref{gauge_large_r})  at orders where supercurrent  components $J_\mu^{(n)}$    are trivial leads to 
\be
 \psi^{(1)}_{0r} = \psi^{(1)}_{1r} = \psi^{(0)}_{1a}= \hat{\gamma}^a\psi^{(0)}_{0a}=  0.
\ee
The  trace-free part  of $\psi^{(0)}_{0a}(u,z)$ is unconstrained and represents free scattering data for gravitino  in the form of two independent function on $\ip$, which correspond to the two independent physical polarizations. The subleading orders of $\psi_\mu$ are expressible in terms of this free data, up to a boundary data.    In particular, the  $\cO(r^{-2})$ order of $\nabla^\mu \psi_\mu=0$  and residual gauge fixing condition (\ref{residual_gauge}) determine
\be\label{scat_u_out}
\psi^{(1)}_{0u}   =   \cD^a \psi^{(0)}_{0a}, 
\ee
while the u-component of RS equation  (\ref{RS_large_r}) 
\be\label{matter_out}
  -\gamma_r \p_u \psi^{(2)}_{1u}  +  \hat{\gamma}^{a}  \cD_a \psi^{(1)}_{0u} =J^{(2)}_{0u}
\ee
determines  $\psi^{(2)}_{1u}(u,z)$ up to its boundary value at $\ip_+$ in terms of the scattering data for gravitino   $\psi^{(0)}_{0a} $ and matter supercurrent $J^{(2)}_{0u}$.

\subsection{Scattering data for gravitino  at $\im$}
Similar analysis can be applied to gravitino field $\psi^-_\mu$ in advanced coordinates  (\ref{adv_coord})  near $\im$ using the same gauge    (\ref{gauge_fix}). Free scattering data at $\im$ is represented in terms of the 
trace free part of the $\psi_{0a}^{(0)-}(u,z)$.  Equations (\ref{scat_u_out}, \ref{matter_out}) in advanced coordinates take the form 
\be\label{matter_in}
\psi^{(1)-}_{0v}   =   -\cD^a \psi^{(0)-}_{0a}, \;\;
  \gamma_r \p_v \psi^{(2)-}_{1v}  +  \hat{\gamma}^{a}  \cD_a \psi^{(1)-}_{0v} =J^{(2)}_{0v}.
\ee

\subsection{Mode expansions}
The mode expansion for the RS field is given by the following formula
  \be\label{modes}
\psi_\mu = \sum_{\a =\pm} \int \frac{d^3q}{(2\pi)^{3}} \;\;[u_\mu^{\ast\alpha} e^{iq\cdot x} b_\a (\vec{q})  + u_\mu^{\alpha} e^{-iq\cdot x}b_\a (\vec{q})^\dagger],
\ee
where  $b_\a(\vec{q})^\dagger$ is a creation operator   for gravitino with spatial momenta $\vec{q}$ and polarization $\a$, that satisfies the commutation relations
\be
\{ b_\a(\vec{p}), b_\b(\vec{q})^\dagger\} =  \delta_{\a\b} (2\pi)^3 \delta^3(\vec{p}-\vec{q}).
\ee  
We can  evaluate the free scattering data $\psi_{0a}^{(0)}$ for RS field in terms of the mode sum using
\be
\psi^{(0)}_{0a}(u,z) = P_0\lim_{r\to \infty} \p_a x^\mu \psi_{\mu}(u+r,r\hat{x}^i(z))= \p_a \hat{x}^i P_0\lim_{r\to \infty} r\psi_i(r,u,z).
\ee
At $r\to \infty$ phases in the mode expansion (\ref{modes}) become large. In particular 
\be
iq\cdot x = -i \omega_q (u+r) +ir q^j \hat{x}_j(z) = -i\omega_q u -ir (1-\hat{x}^j(z)  q_j/\omega_q ),
\ee  
 where $\omega_q $  is the energy for gravitino with spatial momentum $q_i$. The saddle point approximation of (\ref{modes}) leads to 
\be\label{modes_sphere}
\psi^{(0)}_{0a}(u,z) = -\frac{i}{4\pi^2}  \p_a \hat{x}^i(z)  \sum_{\a =\pm} \int^{\infty}_0 \omega_q  d\omega_q \;[u_i^{\ast\alpha} e^{ -i\omega_q u} b_\a (\omega_q \hat{x}(z))  +h.c].
\ee
The positive and negative frequency modes  are given by 
\be  \label{modes2}\begin{split}
\psi_{a}^{\omega (0)}(z)=-\frac{i \omega}{2\pi}\p_a \hat{x}^i (z)\sum_{\alpha}u_{i}^{*\alpha}b_{\alpha}(\omega \hat{x}(z)),
\\
\psi_{0a}^{-\omega ( 0)}(z)=-\frac{i \omega}{2\pi}\p_a \hat{x}^i (z)\sum_{\alpha}u_{i}^{\alpha}b_{\alpha}(\omega \hat{x}(z))^{\dagger}, \end{split}   
\ee
where $\omega >0$ in both formulas. The $\omega \to 0$ limit of these expressions defines a zero mode operator
\be
\psi_{0a}^{0(0)}=\frac{1}{2}\lim_{\omega \to 0}(	\psi_{0a}^{\omega (0)} + \psi_{0a}^{-\omega (0)}		).   
\ee
The asymptotic data at $\mathcal{I}^-$ is given by
\be\label{zero_out} 
\psi_{0a}^{(0)-}(v,z)=P_0 \lim_{r\to \infty}\p_a x^\mu  \psi_{\mu } (v-r,r\hat{x}^i(z)),   
\ee
which may be decomposed into the positive and negative frequency modes
\be \begin{split}
\psi_{0a}^{\omega ( 0)-}(z)=-\frac{i \omega}{2\pi}\p_a \hat{x}^i(z) \sum_{\alpha}u_{i}^{*\alpha}b_{\alpha}(-\omega \hat{x}(z)), 
\\
\psi_{0a}^{-\omega ( 0)-}(z)=-\frac{i \omega}{2\pi}\p_a \hat{x}^i(z)\sum_{\alpha}u_{i}^{\alpha}b_{\alpha}(-\omega \hat{x}(z))^{\dagger}. \end{split}   
\ee
The associated zero mode creation operator is given by
\be\label{zero_in}
\psi_{0a}^{0(0)-}=\frac12\lim_{\omega \to 0}(\psi_{0a}^{\omega ( 0)-}+\psi_{0a}^{-\omega ( 0)-}).   
\ee

\section{Ward identity from soft theorem }
Soft gravitino theorem relates an $n+1$-point on-shell amplitude with a single soft gravitino insertion to a soft operator   acting on $n$-point amplitude. 
Since gravitino is a fermion the soft operator is fermionic as well.  We can add an extra fermionic label to the external states  in addition to physical polarizations and external momenta so that we can consider both bosonic and fermonic states simultaneously.   For massless particles 
external momenta $p^\mu_k$ can be labeled in terms of the energies $E_k$ and  points on a sphere $z_k$, so that 
\be
p^\mu_k  = E_k (1, \hat{x}^i(z_k)).
\ee
For notational simplicity we will suppress all labels except $z_k$ for external incoming and outgoing states 
\be
|z_1,\dots,z_n \>,\;\;\;\; \langle z_{n+1},\dots,z_{n+m}|. 
\ee 
Using an $\mathcal{S}$-matrix presentation for  an $n$-point on-shell amplitude we can write the soft theorem for an outgoing gravitino
\be\label{soft_out}
\lim_{\omega\to 0} \omega \langle z_{n+1}\dots|  b_\a(\vec{q})  \mathcal{S}| z_1\dots\rangle    =  i\left[\sum_{k=n+1}^{n+m} \frac{p_k^\mu (u_{\a\mu} Q_{k})}{p_k\cdot q}
- \sum_{k=1}^n \frac{p_k^\mu (u_{\a\mu} Q_{k})}{p_k\cdot q}\right] \langle z_{n+1}\dots|  \mathcal{S}| z_1\dots\rangle, 
\ee
where $u_{a\mu}$ is a polarization tensor for the gravitino with helicity $\a$; $p^\mu_k$, $Q_{k}$ - momenta and supercharges of matter fields.

Using our results for zero modes of radiation data (\ref{zero_out}) we can write the soft theorem (\ref{soft_out}) in the following from 
\be\label{soft_zero_out}
\langle z_{n+1}\dots|  \psi^{0(0)}_{0a}(z)  \mathcal{S}| z_1\dots\rangle    =- \frac{1}{4\pi}F^{out}_a(z;z_k) \langle z_{n+1}\dots|  \mathcal{S}| z_1\dots\rangle,
\ee
where  functions $F^{out}_a(z;z_k)$\footnote{Note the that the definition of the $F^{out}_a(z;z_k)$ is such that it is proportional  to   expectation value for gravitino's field zero mode  at $\ip$. It would be interesting to design a memory, that can  measure this zero mode. The fermionic nature of zero mode suggests that we may need to couple it to a  superparticle  to create a bosonic observable.}
\be\nn
F^{out}_a(z;z_k) = \omega D_a \hat{x}^i(z)  \sum_\a u_i^{\a\ast} \left[ \sum_{k=1}^{n} \frac{p_k^\mu (u_{\a\mu} Q_{k})}{p_k\cdot q} - \sum_{k=n+1}^{n+m} \frac{p_k^\mu (u_{\a\mu} Q_{k})}{p_k\cdot q}\right]=
\ee
\be
 =\frac{1}{2} \hat{\gamma}_b \hat{\gamma}_a  P_0(z)  \left[ \sum_{k=1}^n Q_k  \p^b \log (1- P(z,z_k) )  -  \sum_{k=n+1}^{n+m} Q_k  \p^b \log (1- P(z,z_k) ) \right].
\ee
Here we used the completeness relation for  polarization tensors 
\be
 \sum_\a u^{\a\ast i} u^{\a j} =(\delta^{ij} -\hat{x}^i \hat{x}^j)P_0  -\frac1{2}\p_a \hat{x}^i \p_b \hat{x}^j  \hat{\gamma}^a \hat{\gamma}^bP_0,
\ee
supercharge conservation 
\be
\sum_{k=1}^n Q_k  -\sum_{k=n+1}^{n+m} Q_k=0, 
\ee
and defined  an invariant  distance on $S^2$ 
\be
 P(z,z_k) \equiv \hat{x}^i(z) \hat{x}_i(z_k).
\ee
Note that $F_a(z;z_k)$ obeys the differential equation 
\be\label{diff_eq}
\sqrt{\gamma}\cD^a F^{out}_a(z;z_k)   = 2\pi \left[\sum_{k=1}^n Q_k \delta^2(z-z_k) -\sum_{k=n+1}^{n+m} Q_k \delta^2(z-z_k)\right]. 
\ee
We can consider a soft theorem for  an incoming soft gravitino
\be\label{soft_in}
\lim_{\omega\to 0} \omega \langle z_{n+1}\dots|   \mathcal{S} b_\a(\vec{q})^\dagger | z_1\dots\rangle    = i\left[\sum_{k=n+1}^{n+m} \frac{p_k^\mu (u_{\a\mu} Q_{k})}{p_k\cdot q}
- \sum_{k=1}^n \frac{p_k^\mu (u_{\a\mu} Q_{k})}{p_k\cdot q}\right] \langle z_{n+1}\dots|  \mathcal{S}| z_1\dots\rangle, 
\ee
and rewrite it using incoming zero mode (\ref{zero_in})
\begin{equation} \label{soft_zero_in}
\langle z_{n+1},\dots|  \mathcal{S}\psi^{0(0)-}_{0a}(z) | z_1,\dots\rangle = \frac{1}{4\pi}F^{in}_{a}(z;z_k) \langle z_{n+1},\dots| \mathcal{S} | z_1,\dots\rangle , 
\end{equation}
where 
\be
F^{in}_{a}(z;z_k)= \frac{1}{2} \hat{\gamma}_b \hat{\gamma}_a  P_0(z)  \left[ \sum_{k=1}^n Q_k  \p^b \log (1+ P(z,z_k) )  -  \sum_{k=n+1}^{n+m} Q_k  \p^b \log (1+ P(z,z_k) ) \right].
\ee

After applying  (\ref{diff_eq}) to equations (\ref{soft_zero_out}) and (\ref{soft_zero_in}), we may integrate against an arbitrary spinor $\bar{\e}(z)$ on the sphere to obtain
\begin{align} \begin{split}
-\int d^2z \sqrt{\gamma}\bar{\e}(z)\gamma^r\cD^a\langle z_{n+1},\dots |\psi_{0a}^{0(0)}(z)\mathcal{S}|z_1,\dots \rangle
+\int d^2z \sqrt{\gamma}\bar{\e}^-(z)\gamma^r\cD^a\langle z_{n+1},\dots |\mathcal{S}\psi_{0a}^{0(0)-}(z)|z_1,\dots \rangle\\
=\left[ \sum_{k=n+1}^{n+m}\e^\dagger(z_k)\cdot Q_k -\sum_{k=1}^{n}\e^\dagger(z_k)\cdot Q_k \right] \langle z_{n+1},\dots |\mathcal{S}|z_1,\dots \rangle,	 \end{split} 
\end{align}
with $\bar{\e} = \e^\dagger \gamma_u$ and $\bar{\e}^-(z)$  being antipodally identified with $\bar{\e}(z)$
what can be written as  Ward identity
\be\label{ward_id}
\langle z_{n+1}\dots |\left(Q^+ \mathcal{S} -\mathcal{S}   Q^-\right)|z_1, \dots \rangle=0.  
\ee
The charges $Q^\pm= Q_H^\pm + Q_S^{\pm}$ commute with the $\mathcal{S}$-matrix and induce infinitesimal symmetry transformations on   $\mathcal{I}^{\pm}$ states. $Q_H^{\pm}$ is defined by its action on the asymptotic states: 
\be \label{hard_ch}
Q^-_H|z_1,\dots \rangle = \sum_{k=1}^n \e^\dagger (z_k)\cdot Q_k  |z_1,\dots\rangle,\;\;\  \langle z_{n+1},\dots |Q_{H}^+ =   \langle z_{n+1},\dots| \sum_{k=n+1}^{n+m} \e^\dagger(z_k)\cdot Q_k.   
\ee 
The soft charges are given by
\be\label{soft_ch_out}
Q^+_S =    \int  d^{2}z\; \bar{\e} \gamma^r \cD^a\psi^{0(0)}_{0a}  ,
\ee
\be\label{soft_ch_in}
Q^-_S =   \int  d^{2}z\;  \bar{\e}^- \gamma^r \cD^a\psi^{0(0)-}_{0a}  . 
\ee

\section{From Ward identity to asymptotic symmetry}

\subsection{Action of matter charges}
The hard charges (\ref{hard_ch}) defined by their action on the in- and out- states can be written using super current 
\be
Q^+_H = \lim_{r\to \infty}  r^2\int_{\ip}  \sqrt{\gamma} d^2z du\;\; \bar{\e} J_{0u}  (u,r,z) 
\ee
\be
Q^-_H = \lim_{r\to \infty}  r^2\int_{\mathcal{I}^-}  \sqrt{\gamma} d^2z dv\;\; \bar{\e}^{-} J_{0v}  (v,r,z) 
\ee
These expressions can be further rewritten in the form 
\be\label{cgen}
Q_{H}^+=\lim_{\Sigma\to \ip}  \int_{\Sigma} d\Sigma\; n^\mu_{\Sigma}  \bar{\e}  J_{\mu},\;\;\;\;\;\;\;\;\;\;\; Q_{H}^-=\lim_{\Sigma\to \im}  \int_{\Sigma} d\Sigma\; n^\mu_{\Sigma}  \bar{\e} J_{\mu}.  
\ee
Here $\Sigma$ is a space-like Cauchy surface, $n_\Sigma$ is a unit normal to $\Sigma$. Written in this form, it is clear that the hard charges generate supersymmetry transformations with parameter $\e$  on the asymptotic states if we assume  standard commutation relations for the matter fields.

\subsection{Vanishing soft charge}
Symmetry charges (\ref{hard_ch}, \ref{soft_ch_in}, \ref{soft_ch_out})  apply a supersymmetry transformation to matter fields and create a zero mode for gravitino. However global supersymmetries are just a symmetries of the matter sector of the coupled theory. Thus  the spinors $\e$ for which the soft charge   (\ref{soft_ch_out}) vanishes should represent  global supersymmetries of Minkowski space.
Since $\psi^{0(0)}_{0a}$ is trace-free the most general such $\e$ is a solution to 
\be
\cD_a \e = \hat{\gamma}_a \omega
\ee
 for some spinor $\omega$.  Using (\ref{covariant_com}) we can show that 
\be
\cD_a \omega = -\frac14  \hat{\gamma}_a \e.
\ee
The linear combinations 
\be
\e_{\pm}= \e \pm 2\gamma^r \omega
\ee
solve the canonical Killing spinor equation on the two-sphere
\be\label{killing_spinor}
\cD_a \e_{\pm} = \pm\frac 12 \gamma^r \hat{\gamma}_a \e_{\pm}.
\ee
Furthermore $\e^{\pm}(z)$ trivially satisfies 
\be
\p_u \e_{\pm}= \p_r \e_{\pm}=0, 
\ee
what allows us to use $\e^{\pm}(z)$ to parametrize   solutions to covariantly constant spinor in Minkowski.

\subsection{Action of soft charges}
Let us use our prediction for the symplectic structure in conjugation with our proposal for  the soft charge. From the mode expansion (\ref{modes_sphere}) we can read off the following bracket
\be
\{\psi_{a}^{(0)}(u,z), \psi_{b}^{(0)\dagger}(v,w)\} =\left(\gamma_{ab} - \frac12 \hat{\gamma}_a\hat{\gamma}_b\right) P_0 \delta^{(2)}(z-w) \delta(u-v),
\ee
and use it to evaluate the soft charge action
\be
\delta_\e \psi^{(0)}_{a}= \left(\gamma_{ab} - \frac12 \hat{\gamma}_a\hat{\gamma}_b\right) \cD^b \e_0  = \cD_a \e_0 - \frac12 \hat{\gamma}_a \hat{\gamma}^b \cD_b \e_0.
\ee
The  local gauge transformation (\ref{gauge_transform}) of $\psi_{0a}^{(0)}$
\be
\delta \psi_{0a}^{(0)} =  \cD_a \e_0  - \frac12 \gamma^r \hat{\gamma}_a \e_1
\ee
contains residual gauge transformation that preserves (\ref{gauge_large_r}), (\ref{residual_gauge}) for 
\be\label{Dirac_solution}
\e_1 = -\gamma^r \hat{\gamma}^a \cD_a \e_0. 
\ee
Therefore we conclude that $Q^+_S$ acts linearly on zero mode of gravitino 
\be\label{goldstone_out}
 \psi_{0a}^{(0)}(z,u) =  \cD_a \sigma_0(z) - \frac12 \hat{\gamma}_a \hat{\gamma}^b \cD_b \sigma_0(z),\;\;\;\; \delta_\e \sigma_0(z)=\e_0(z). 
\ee
Similarly for the scattering data on $\im$ we have 
\be\label{goldstone_in}
 \xi_{0a}^{(0)}(z,v) =  \cD_a \sigma^-_0(z) - \frac12 \hat{\gamma}_a \hat{\gamma}^b \cD_b \sigma^-_0(z),\;\;\;\; \delta_{\e^-} \sigma^-_0(z)=\e_0^-(z). 
\ee

 \subsection{Scattering}
So far we treated  $\ip$ and $\im$ separately and symmetry charges $Q^{\pm}$ acted on two separate scattering datasets.  In order to connect $\mathcal{I}^-$ to $\mathcal{I}^+$ we must match the  scattering  data at $i^0$. Following the analysis in \cite{Strominger:2013jfa}, all fields and functions are taken to be continuous along the null generators of $\mathcal{I}$ passing through $i^0$. Due to the antipodal identification of the angular coordinates on $\mathcal{I}^+$ and $\mathcal{I}^-$, the zero modes (\ref{goldstone_out}) for  $\psi_{0a}^{(0)}$ and (\ref{goldstone_in}) for   $\psi_{0a}^{(0)-}$ are matched according to
\be
\sigma_0(z)=\sigma^-_0(z).   
\ee
This identification also allows for a canonical identification of transformation parameters  according to the rule
\be
\e(z)=\e^{-}(z),   
\ee
yielding a diagonal  subgroup that may be identified as a symmetry of the  $\mathcal{S}$-matrix.

\subsection{Total charges}
Typical total charge in theory with a gravity is total derivative on corresponding part of null infinity. In our case we have the following expression for the total charge
\be
Q^+ = Q_S^++ Q_{H}^+ =  \int  du d^2z \sqrt{\gamma}\; \bar{\e} (  \gamma^r\cD^a\psi^{(0)}_{0a}+ J^{(2)}_{0u} ).
\ee
Using (\ref{Dirac_solution}) and (\ref{matter_out})  we can write total charge in the form 
\be\label{}
Q^+ =   -  \int_{\ip}   \bar{\e}  \gamma_r \p_u \psi^{(2)}_{u1}. 
\ee 
Furthermore  we can impose an additional boundary condition 
\be
 \psi^{(2)}_{u1}|_{\ip_+}=0,
\ee
so that  the total charge is an integral over $\ip_-$ surface, which we use to match the scattering data.     This charge is similar to the charge introduced for global super symmetries in\cite{Horowitz:1982mu}.

 \section{Concluding remarks}
 At the last stages of the work we were contacted by the Steven Avery and Burkhard U.W. Schwab \cite{Avery}, who  work on a  similar problem.  Authors take a different approach to the analysis  of  asymptotic supersymmetries in $\mathcal{N}=1, d=4$ supergravity, while the final conclusion is similar to ours: There is an infinite dimensional fermonic symmetry for the garavitino coupled to mater in four dimensional Minkowski space. 
 
 It is worth mentioning that we used notations which can be adapted to the higher dimensional cases, in a similar way as we did in  \cite{Kapec:2014zla,Kapec:2015vwa}. The whole construction of the new symmetries from soft gravitino theorem might as well work in higher dimensions.

\section{Acknowledgments}
I am  grateful to  for  T. He and P. Mitra for their comments on the draft of the  article.  I am   grateful  to T. Dumitrescu, S. Gukov,   J. Schwarz and A. Strominger  for useful conversations. 
 I  also thank participants and organizers  of the``Quantum Gravity Foundations: UV to IR" program at Kavli Institute for Theoretical Physics  for useful discussions. My work  is supported in part by DOE grant DE-SC0011632 and  the Sherman Fairchild scholarship.

\end{document}